# Structural, Dielectric and Ferroelectric Properties of Lead-free BCZT Ceramics Elaborated by Low-temperature Hydrothermal Processing

Zouhair Hanani [a,b,*], Daoud Mezzane [a], M'barek Amjoud [a], Yaovi Gagou [c], Khalid Hoummada [d], Carine Perrin [d], Anna G. Razumnaya [c,e], Zdravko Kutnjak [f], Adnane Bouzina [a], Mimoun El Marssi [c], Mohamed Gouné [b] and Brigita Rožič [f]

[a] LMCN, Cadi Ayyad University, Marrakesh, 40000, Morocco
[b] ICMCB, University of Bordeaux, Pessac, 33600, France
[c] LPMC, University of Picardy Jules Verne, Amiens, 80039, France
[d] IM2NP, Aix Marseille University, Marseille, 13397, France
[e] Faculty of Physics, Southern Federal University, Rostov-on-Don, 344090, Russia
[f] Jožef Stefan Institute, Ljubljana, 1000, Slovenia

[*] *Corresponding author:* zouhair.hanani@edu.uca.ma

## Abstract

Lead-free $Ba_{0.85}Ca_{0.15}Zr_{0.10}Ti_{0.90}O_3$ (BCZT) ceramics have demonstrated excellent dielectric, ferroelectric and piezoelectric properties in comparison to lead-based materials. The synthesis of pure and crystalline BCZT nanopowders at low temperatures of 25, 80 and 160 °C was reported previously by using a sol-gel method followed by a hydrothermal route [1]. In this study, the structural, dielectric and ferroelectric properties of sintered BCZT ceramics at 1250 °C/10h were investigated. XRD measurements revealed the presence of a single perovskite phase at room temperature with the coexistence of the orthorhombic and tetragonal symmetries. The increase of grain size and the ceramic density in BCZT ceramics result in an enhancement of the dielectric and ferroelectric properties of BCZT ceramics. More interestingly, the synthesis temperature of BCZT powders with high dielectric and ferroelectric properties could be decreased to a low temperature of 160 °C, which is about 1200 °C lower when compared with solid-state reaction and 840 °C lower when compared with sol-gel methods, respectively. The BCZT ceramics elaborated at 160 °C revealed excellent electrical properties ($\varepsilon_m$ = 12085, $tan\ \delta$ = 0.017, $P_r$ = 8.59 μC/cm² and $P_{max}$ = 27.21 μC/cm²). Hence, the use of low-temperature hydrothermal processing can be encouraging for the synthesis of lead-free ceramics with high dielectric and ferroelectric properties.

# 1. Introduction

Lead-based materials like $PbZr_xTi_{1-x}O_3$ (PZT) have been thoroughly investigated in last decades due to their outstanding dielectric, ferroelectric and piezoelectric properties for application in energy storage devices, capacitors, transducers and sensors [2,3]. However, their possible toxicity limits their applications in future. Correspondingly, new lead-free materials were developed recently to mimic properties of lead-based materials [4–6].

Since its investigation by Liu and Ren [7], $Ba_{0.85}Ca_{0.15}Zr_{0.10}Ti_{0.90}O_3$ (BCZT) ceramic was mainly explored due to its excellent dielectric and electromechanical properties ($\varepsilon_r \approx 18\,000$ and $d_{33} \approx 620$ pC/N) at the morphotropic phase boundary (MPB) that are comparable to lead-based materials. Subsequently, versatile works were conducted to thoroughly understand the relationship of microstructural-electrical properties to tailor BCZT ceramic with outstanding dielectric, ferroelectric and piezoelectric properties [8–11]. Bai et al. [12] reported the critical role of $Ba_{1-x}Ca_xZr_yTi_{1-y}O_3$ processing on their properties, including $Ba_{0.85}Ca_{0.15}Zr_{0.10}Ti_{0.90}O_3$ (BCZT) ceramic. They investigated the effect of a wide range of processing factors, including sintering conditions (temperature and cooling rate), particle size of the calcined ceramic powder, structure and microstructure, and their influence on the electrical properties.

Energy consumption reduction is one of crucial aspects for industrial application of lead-free BCZT ceramic. However, a large and growing body of literature is based on high temperature synthesis to make pure and crystalline BCZT powder [8,13,14]. The first published synthesis of BCZT complex composition was based on solid-state reaction at high temperatures (calcination at 1350 °C and sintering between 1450 – 1500 °C) due to its simplicity and capability to obtain enhanced dielectric, ferroelectric and piezoelectric performances [7]. Afterwards, BCZT ceramic was elaborated by soft chemistry routes like sol-gel [13], citrate reaction [15], solvothermal [11] and hydrothermal [16] method. However, here high temperature is needed to obtain pure and crystalline BCZT powder. Jaimeewong et al. [17] outlined a comparative study of two BCZT ceramics synthesized by solid-state and sol-gel auto combustion methods at 1200 °C and 900 °C, respectively. Quite recently, we reported the effect of grain size on the dielectric properties of BCZT ceramic via surfactant-assisted solvothermal processing at low-temperature sintering [11]. Enhanced dielectric and ferroelectric properties were obtained using the anionic surfactant SDS (sodium dodecyl sulfate). However, the high calcination temperature (1000 °C) limits the application of BCZT ceramic in the energy-efficient industrial fields.

To overcome the drawback of energy consumption during calcination, we raised a challenge to elaborate pure and crystalline BCZT ceramic at very-low temperature without calcination step.

Previously, we reported the synthesis of scalable BCZT ceramics at low temperature including room-temperature (25 °C) [1]. The key strategy is to elaborate BCZT ceramics using two-steps synthesis to overcome the low reactivity of Zr ions [18,19]. First, sol-gel to obtain $Zr_{0.10}Ti_{0.90}O_2$, then addition of Barium and Calcium salts to obtain pure BCZT crystalline nanopowders by hydrothermal route. More recently, Ji et al. [20] reported the effects of synthesis time on the structure, morphology and size distribution of the powders of BCZT ceramic elaborated by sol-gel−hydrothermal reaction at 180 °C. However, high sintering-temperature (1400 °C) was required to obtain enhanced dielectric, ferroelectric and piezoelectric properties. In this study, we aim to evaluate the structural, morphological, dielectric and ferroelectric properties of BCZT ceramic sintered at low sintering temperature (1250 °C) using powders elaborated at very-low temperature (25, 80 and 160 °C) [1].

## 2. Experimental

*2.1. BCZT Powders synthesis*

BCZT pure and scalable nanocrystalline powders were obtained through sol-gel method followed by a single-step hydrothermal synthesis in high alkaline medium as reported previously [1]. First, B-site (ZT) of BCZT ceramic was elaborated via sol-gel method by reacting stoichiometric amounts of titanium $^{(IV)}$ isopropoxide and zirconium n-propoxide in isopropanol under inert atmosphere ($N_2$). Further, ZTO powder was obtained by dropwise addition of distillated water into the solution to produce a gel designated ZTO, washing it several times with distilled water and ethanol and drying the gel. Next, a stoichiometric quantity of ZTO powder was dispersed in NaOH (10 M) solution. Second, A-site (BC) of BCZT was obtained by dissolving calcium nitrate tetrahydrate and barium acetate distillated water. Then, the two solutions were mixed under nitrogen flow. The obtained suspension was transferred to a Teflon-lined stainless-steel autoclave, purged with nitrogen, sealed and heated at 25, 80 and 160 °C for 24 h and designated as B-25, B-80 and B-160, respectively. After the reaction was completed, the sealed autoclave was cooled down to the room temperature. The resulting white precipitates were extracted by centrifugation, purified and dried. It is worth noting that no additional thermal treatment like calcination was performed. For electrical measurements, BCZT powders were uni-axially pressed into pellets of diameter about 6 mm and thickness about 1 mm, and then sintered at 1250 °C/10h.

*2.2. Characterization*

Crystalline structure of BCZT ceramics was determined by X-ray diffraction (XRD, Rigaku) using a Cu-K$_\alpha$ radiation ($\lambda \sim 1.540593$Å). The resulting microstructures were analyzed by using a Scanning Electron Microscope (SEM, Tescan VEGA3). The grain size distributions of BCZT ceramics were determined by using ImageJ software. The density of the sintered ceramics was measured by Archimedes method using deionized water as a medium. A precision LCR Meter (HP 4284A, 20 Hz to 1 MHz) was used to measure the dielectric properties of gold-sputtered BCZT pellets in the frequency range of 20 Hz to 1 MHz. The ferroelectric hysteresis loops were determined by using a ferroelectric test system (AiXACCT, TF Analyzer 3000).

## 3. Results and discussion

### 3.1. Morphological study

Fig. 1 displays the SEM micrographs and the variation of bulk and relative densities and grain size with synthesis temperature for BCZT ceramics sintered at 1250 °C/10h. B-25 exhibits small grain and monomodal grain distribution (Fig. 1a). However, B-80 shows bimodal grain distribution with the presence of small and coarse spheroidal grains (Fig. 1b). Contrary to B-25 and B-80, B-160 ceramic exhibits a well-defined grain boundary with coarse and flatten grains leading to an enhancement of bulk and relative densities (Fig. 1d). In B-160 ceramic the bulk density was found to be 5.62 g/cm$^3$, which corresponds to 97.1% of the theoretical density (5.789 g/cm$^3$).

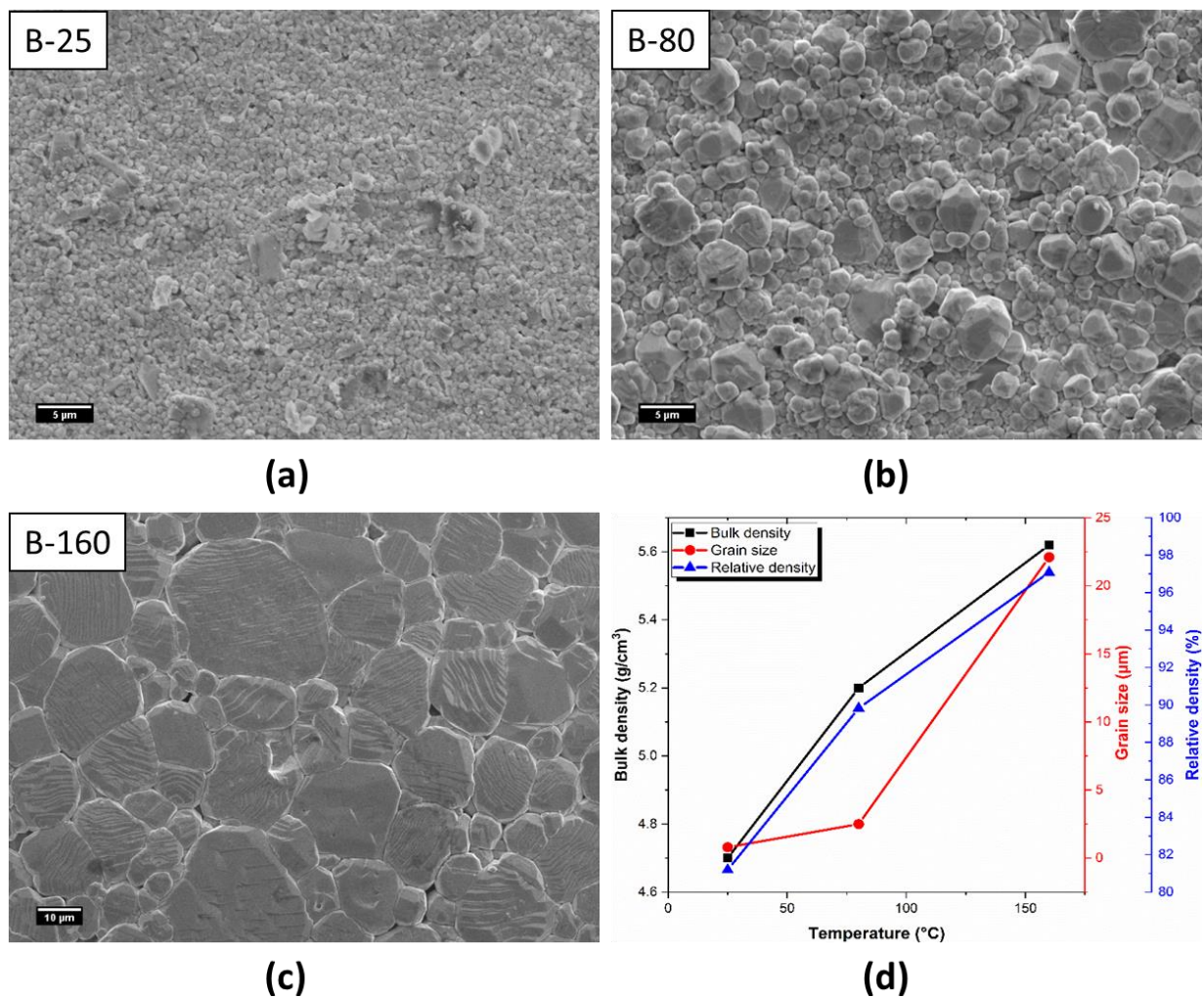

**Fig. 1** SEM micrographs of (a) B-25, (b) B-80 and (c) B-160 ceramics sintered at 1250 °C/10h and (d) Variation of bulk and relative densities and grain size with synthesis temperature

*3.2. Structural study*

Fig. 2 reveals the XRD patterns of BCZT ceramics elaborated at different temperatures. As we reported previously in the as-prepared BCZT powders [1], all the sintered BCZT ceramics have a pure perovskite phase, without any secondary phase peaks, which indicates that the stabilization of the ferroelectric phases was already reached in BCZT powders. The diffraction peaks of all samples shift to higher *2θ* values by increasing the hydrothermal temperature (Fig. 2b). In other words, diffraction peaks shift to higher diffraction angles as grain size increases as seen in SEM images in Fig. 1. These suggest a decreasing of BCZT unit cell, and/or the reduction of defects and the structural relaxation [21,22]. Meanwhile, the splitting of peaks observed around 45° (Fig. 2c) and the formation of triplets at 65.8° (Fig. 2d) demonstrate the coexistence of orthorhombic (*Amm2*) and tetragonal phases (*P4mm*) at the morphotropic phase boundary (MPB) in BCZT ceramics [1,11,23–25].

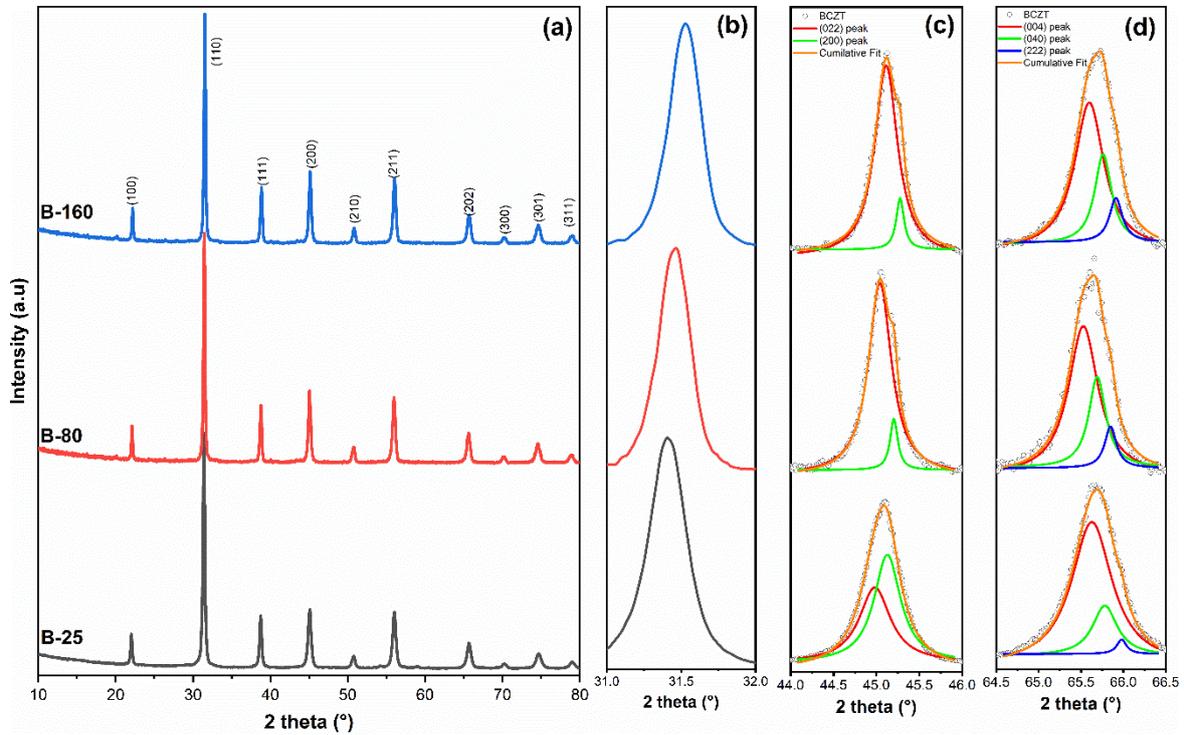

**Fig. 2** (a) XRD patterns, (b) enlarged peak at *2θ* ≈ 31.5° and peak splitting around (c) *2θ* ≈ 45° and (d) *2θ* ≈ 65.8° of B-25, B-80 and B-160 ceramics sintered at 1250 °C/10h

*3.3. Dielectric properties*

Fig. 3a compares the temperature-dependence of the dielectric constant and dielectric loss at 1 kHz of B-25, B-80 and B-160 ceramics sintered at 1250 °C/10h. All samples exhibit a broad dielectric anomaly associated with tetragonal-cubic (*T–C*) phase transition (Fig. 3a). Meanwhile, the dielectric losses of all samples decrease gradually to 40 °C, then remain constant and after they increase (Inset of Fig. 3a). The overall dielectric properties at 1 kHz of all ceramics are summarized in Table 1. The temperature dependence of the dielectric constant and the dielectric loss of all BCZT ceramics at various frequencies are shown in Figs. 3b - 3d. The dielectric properties were enhanced with increasing hydrothermal temperature. Despite the low sintering temperature of B-160 ceramic, the maximal value of the dielectric constant reached a value of 12085, which is much higher than that reported in other works using solvothermal [11], hydrothermal [16] or sol-gel hydrothermal routes [20]. Moreover, for all BCZT samples the peak temperatures are frequency-dependent and shifted to higher temperature with increasing frequency (reminiscent of relaxor behavior) [26,27].

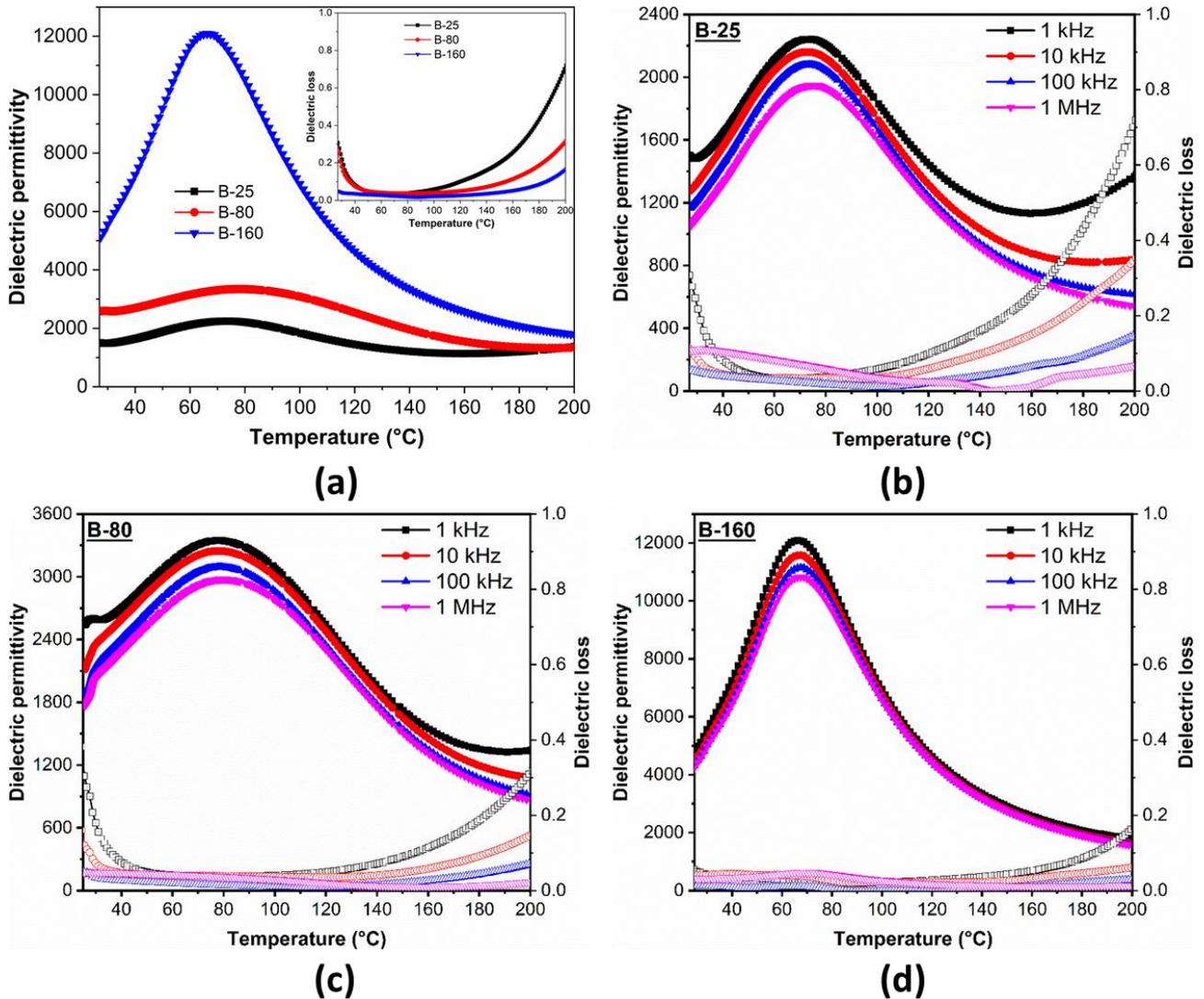

**Fig. 3** (a) Comparison of the dielectric properties of BCZT ceramics at 1 kHz, and temperature dependence of the dielectric permittivity and dielectric loss of (a) B-25, (b) B-80 and (c) B-160 ceramics sintered at 1250 °C/10h

To have an insight into this phase transition, the inverse dielectric constant was plotted as a function of temperature at 1 kHz and fitted by using the Curie–Weiss law:

$$\frac{1}{\varepsilon_r} = \frac{T - T_0}{C} (T > T_0), \qquad (1)$$

where $\varepsilon_r$ is the real part of the dielectric constant, $T_0$ is the Curie Weiss temperature and $C$ is the Curie-Weiss constant.

The inverse of the dielectric constant as a function of temperature at 1 kHz for all BCZT ceramics is plotted in Fig. 4(a, c, e), and the fitting results obtained by using Eq. (1) are listed in the Table 1. It should be noted that the dielectric constant ($\varepsilon_r$) of all samples deviate from the Curie–Weiss law above the Curie temperature. Eq. (2) evaluates the this deviations by introducing $\Delta T_m$ that is defined as [28],

$$\Delta T_m(K) = T_{dev} - T_m, \quad (2)$$

Here, $T_{dev}$ designates the temperature below which $\varepsilon_r$ starts to deviate from the Curie–Weiss law, and $T_m$ denotes the temperature at which $\varepsilon_r$ value reaches the maximum. The calculated $\Delta T_m$ of B-25, B-80, and B-160 ceramics are shown in Table 1. The highest value was found to be 37 °C for B-80. However, the lowest $\Delta T_m$ was determined in B-25 ceramic.

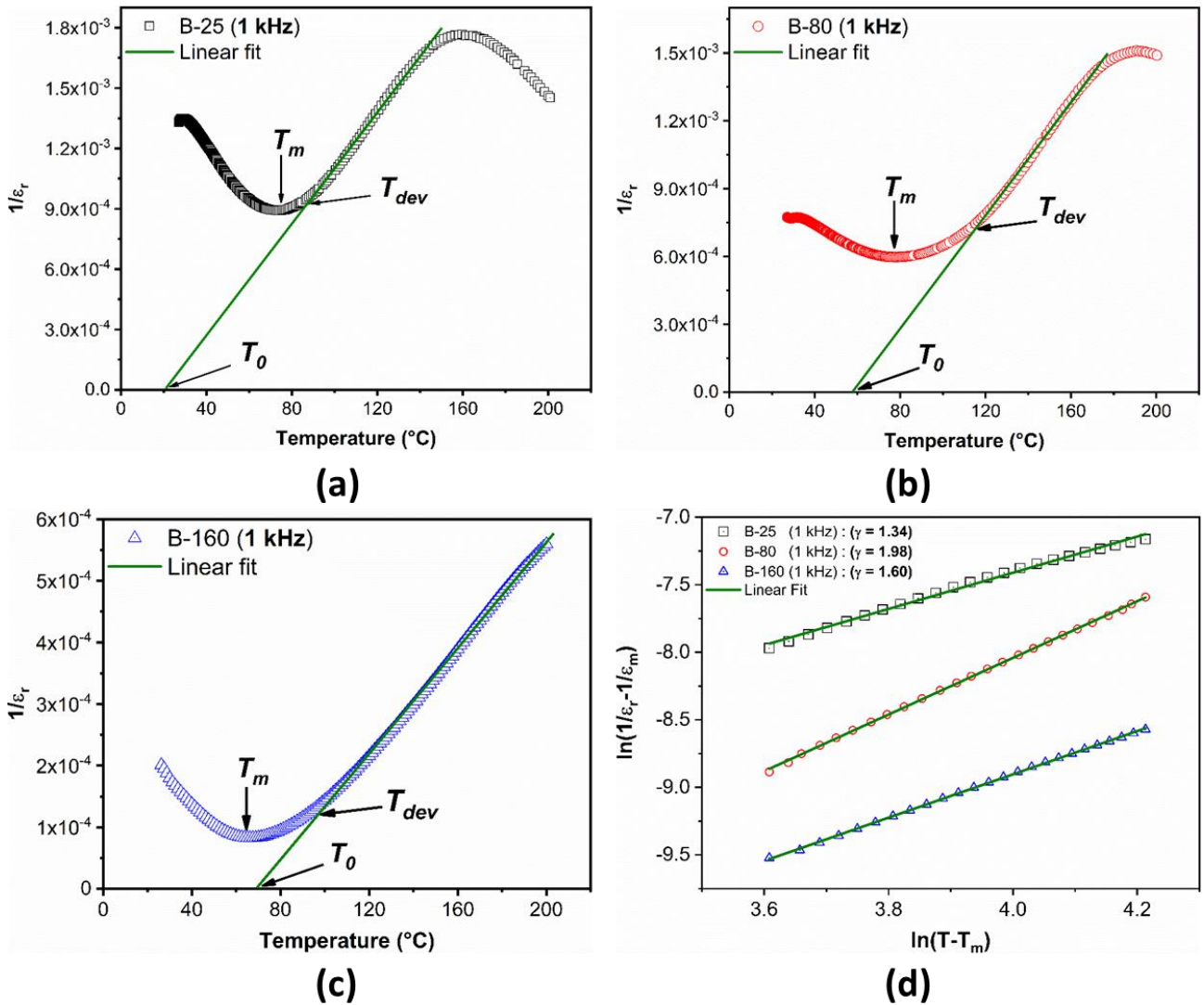

**Fig. 4** Plots of Curie-Weiss relation for (a) B-25, (b) B-80 and (c) B-160 ceramics sintered at 1250 °C/10h, and (d) modified Curie-Weiss law to determine slope ($\gamma$)

All ceramics are found to be associated with the Curie constant value of the order of $10^5$ K, which are consistent with that of typical well known displacive-type ferroelectrics such as BaTiO$_3$ (1.7×10$^5$ K) [29]. This indicates that a high-temperature ferroelectric phase is driven by a displacive phase transition. To clarify the diffuseness associated with the Tetragonal-Cubic transition in general, the Uchino and Nomura [30] empirical relation is used to describe the variation of dielectric constant as a function of temperature above $T_c$ for relaxors. The relation is defined as follows [30]:

$$\frac{1}{\varepsilon_r} - \frac{1}{\varepsilon_m} = \frac{(T-T_0)^\gamma}{C} \quad (1 < \gamma < 2), \qquad (3)$$

where $\varepsilon_r$, $\varepsilon_m$ are the real part of the dielectric permittivity and its maximum value, respectively, and $\gamma$ (degree of diffuse transitions) and $C$ are constants. For an ideal relaxor ferroelectric $\gamma$ =2, while for a normal ferroelectric $\gamma$ =1 and the system follows the Curie-Weiss law [16]. Fig. 4 (b, d, f) depict the linear relationship between $ln\ (1/\varepsilon_r - 1/\varepsilon_m)$ and $ln\ (T - T_m)$ at 1 kHz for all BCZT ceramics. After curves fitting by using Eq. (3), the values obtained for $\gamma$ at 1 kHz in B-25, B-80 and B-160 ceramics are 1.34, 1.98 and 1.60, respectively. The high $\gamma$ values in B-25 and B-80 ceramics indicate that the diffuse phase transition is supported by the small grain size [16,31]. In this context, enhanced relaxor behavior is obtained while decreasing the grain size in BCZT ceramics [32,33].

|  | $\varepsilon_m$ | tan δ | Grain size (µm) | $T_0$ (°C) | $C \times 10^5$ (K) | $T_m$ (°C) | $T_{dev}$ (°C) | $\Delta T_m$ (°C) | $\gamma$ |
|---|---|---|---|---|---|---|---|---|---|
| **B-25** | 2242 | 0.035 | 0.8 | 21 | 0.72 | 74 | 87 | 13 | 1.34 |
| **B-80** | 3346 | 0.037 | 2.5 | 58 | 0.80 | 78 | 115 | 37 | 1.98 |
| **B-160** | 12085 | 0.017 | 22.1 | 71 | 2.33 | 66 | 95 | 29 | 1.60 |

**Table 1** Relaxor properties at 1 kHz of B-25, B-80 and B-160 ceramics sintered at 1250 °C/10h

*3.4. Ferroelectric properties*

To have an insight on the ferroelectric properties of BCZT ceramics, room-temperature *P-E* loops under 20 kV/cm and 10 Hz were plotted in Fig. 5a. Interestingly, it was demonstrated for the first time, that the ceramic elaborated at room temperature (B-25) exhibits a ferroelectric behavior evidenced by the P-E hysteresis loops. The remnant polarization increases from 0.81 to 2.80 and 4.52 µC/cm² for B-25, B-80 and B-160, respectively. In contrast, the coercive field decreases from 3.63 to 3.97 and 3.21 kV/cm for B-25, B-80 and B-160, respectively. Thus, increasing the synthesis temperature (grain size increasing) results in an enhancement of the ferroelectric properties. Interestingly, the non-saturated *P-E* loop in B-160 ceramics could help to further increase in the applied electric field. Fig. 5b depicts the hysteresis loops of B-160 ceramic at various electric fields from 10 to 60 kV/cm. It is shown that B-160 ceramic exhibits enhanced ferroelectric properties, the maximal polarization ($P_{max}$) and remnant polarization ($P_r$) increase with the increasing of the applied electric field. Under 60 kV/cm, $P_{max}$ and $P_r$ achieved 27.21 µC/cm², 8.59 µC/cm², respectively. The excellent ferroelectric properties in B-160 ceramics further confirms the advantages of the low-temperature hydrothermal processing.

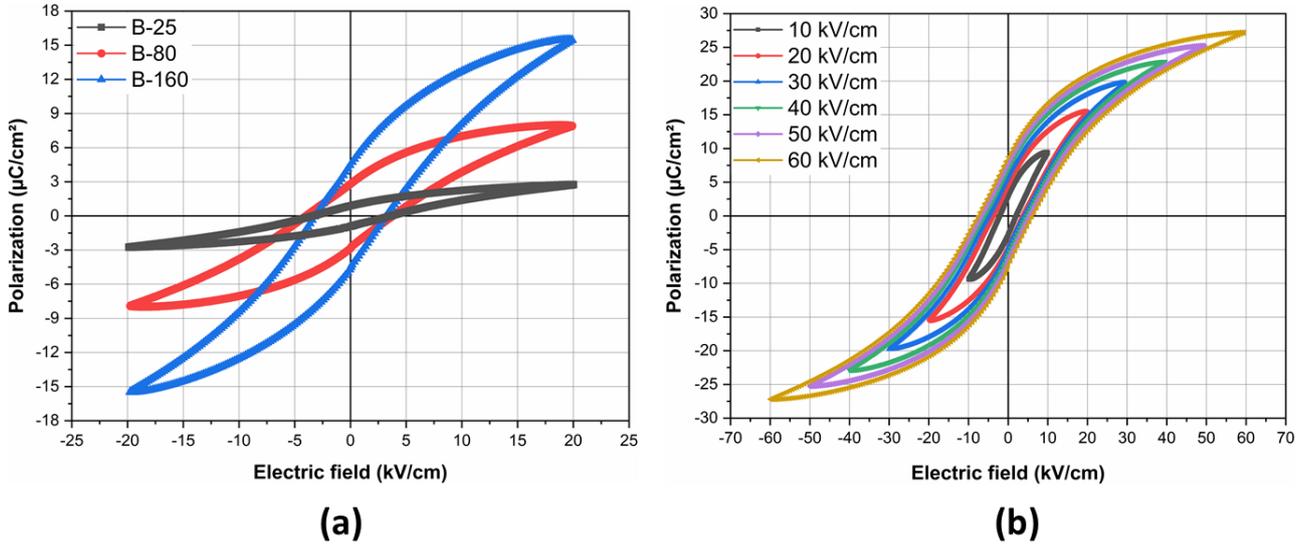

**Fig. 5** (a) Room-temperature *P-E* loops of BCZT ceramics under 20 kV/m and (b) *P-E* loops of B-160 ceramic under various applied electric fields at 10 Hz

To illustrate the advantages of the low-temperature hydrothermal processing in lead-free BCZT ceramics elaboration, Table 2 summarizes the grain size average, relative density, dielectric and ferroelectric properties of $Ba_{0.85}Ca_{0.15}Ti_{0.9}Zr_{0.1}O_3$ (BCZT) ceramic elaborated by different conditions of synthesis and methods. From energy consumption view, the synthesis temperature of BCZT powders with high dielectric and ferroelectric properties could be decreased to a low temperature of 160 °C, which is about 1000 − 1200 °C lower when compared with solid-state reaction and 640 − 840 °C lower when compared with sol-gel methods respectively. Among the big advantage of "calcination-free" in hydrothermal and sol-gel-hydrothermal processing, these strategies allow the synthesis of BCZT ceramics with enhanced dielectric and ferroelectric properties. For instance, B-160 ceramic processes a maximal dielectric constant of 12085 higher than BCZT ceramics elaborated by solid-state and sol-gel−hydrothermal processing, while keeping low dielectric loss. Moreover, from energy storage consideration, according to the following equations, large dielectric breakdown strength, a reduced remnant polarization and large maximal polarization will lead to improved recoverable energy density and energy storage efficiency [34]. In other words, the charge storage density ($Q_c = P_{max} - P_r$) in a ferroelectric capacitor calculated from the ferroelectric hysteresis loops at zero field, must be maximal to obtain high energy storage ferroelectric capacitor [35]. These requirements are gathered in BCZT ceramics elaborated by hydrothermal processing. B-160 could withstood high electric field (60 kV/cm) and shows $Q_c$ = 18.62 µC/cm². Hence, low-temperature hydrothermal route could be suitable for the design of energy harvesting systems with improved performances.

$$W = \int_0^{P_{max}} E\, dP, \quad (4)$$

$$W_{rec} = \int_{P_r}^{P_{max}} E\, dP, \quad (5)$$

$$\eta(\%) = \frac{W_{rec}}{W} \times 100, \quad (6)$$

where $W$, $W_{rec}$ and $\eta$ are charged energy density, recoverable energy density and energy efficiency, while $P_{max}$, $P_r$, $E$ and $P$ represent maximum polarization, remnant polarization, electric field and polarization, respectively.

| Synthesis conditions | | | Grain size | Relative density | $\varepsilon_m$ | tan $\delta$ | $P_r$ | $P_{max}$ | $Q_c$ | $E$ | Ref. |
|---|---|---|---|---|---|---|---|---|---|---|---|
| Method | Calcination | Sintering | (µm) | (%) | | | (µC/cm²) | (µC/cm²) | (µC/cm²) | kV/cm | |
| Hydrothermal | No | 1250 °C/10h | 22.1 | 97.1 | 12085 | 0.017 | 8.59 | 27.21 | 18.62 | 60 | This work |
| Sol-gel−hydrothermal | No | 1400 °C/2h | - | 95 | 9173 | - | 12.56 | 41.00 | 28.44 | 40 | [20] |
| Hydrothermal | No | 1300 °C/3h | 12.09 | - | 7760 | 0.1 | 10.83 | 25.00 | 14.17 | 15 | [16] |
| Solvothermal | 1000 °C/4h | 1250 °C/10h | 6.6 | 96.4 | 9646 | 0.012 | 3.92 | 7.56 | 3.64 | 6.6 | [25] |
| Sol-gel | 1000 °C/4h | 1420 °C/6h | - | - | 16480 | 0.015 | 11.60 | 17.76 | 6.16 | 30 | [13] |
| Sol-gel | 800 °C | 1400 °C/2h | - | 95 | 8808 | 0.02 | 12.20 | 17.23 | 5.03 | 30 | [36] |
| Sol-gel | 1000 °C/4h | 1550 °C/2h | 10 | 97 | 20250 | - | 10.70 | 20.70 | 10.00 | 50 | [37] |
| Sol-gel auto-combustion | 1200 °C/2h | 1450 °C/2h | 3.4 | 97 | - | - | 7.38 | 15.50 | 8.12 | 20 | [17] |
| Solid-state | 1150 °C/6h | 1400 °C/6h | - | 95 | 10615 | - | 8.21 | 18.53 | 10.32 | 50 | [38] |
| Solid-state | 1147 °C/12h | 1427 °C/2h | - | - | 4762 | 0.022 | 4.35 | 6.53 | 2.18 | 8 | [39] |
| Solid-state | 1350 °C and 1400 °C/6h | 1450 °C/4h | - | 94 | - | - | 5.48 | 16.06 | 10.58 | 21 | [40] |

**Table 2** Comparison of the dielectric and ferroelectric properties of lead-free BCZT ceramics reported here with others reported in literature using different synthesis conditions and methods

## 4. Conclusions

Structural, morphological, dielectric and ferroelectric properties of three lead-free BCZT ceramics (B-25, B-80 and B-160) prepared via hydrothermal route at very-low temperature (≤ 160 °C) have been investigated. X-ray diffraction demonstrated that all BCZT samples were crystallized in a pure perovskite phase with the coexistence of orthorhombic/tetragonal structure at room temperature. Increasing the hydrothermal treatment, i.e. increasing grain size, enhanced the dielectric and ferroelectric properties of BCZT ceramics. In B-160 ceramic, the ceramic density reached 97.1%, and the dielectric constant and dielectric loss values at the maximum of the temperature peak of the dielectric constant were of 12085 and 0.017, respectively. Moreover, the low $P_r$ of 8.59 µC/cm² and the high $P_{max}$ of 27.21 µC/cm², could be encouraging for the integration of BCZT ceramics in energy harvesting systems.

## Acknowledgements

The authors gratefully acknowledge the generous financial support of CNRST Priority Program PPR 15/2015 and the European Union's Horizon 2020 research and innovation program under the Marie Skłodowska-Curie grant agreement No. 778072. Z. K. and B. R. acknowledge Slovenian Research Agency grant J1-9147 and program P1-0125.